\documentclass{ws-procs975x65}            

\begin{document}
\title{Transpose-free Fast Fourier Transform for Turbulence Simulation}

\author{A. G. Chatterjee and M. K. Verma$^*$}

\address{Department of Physics, IIT Kanpur,\\
Kanpur, 208016, India\\
$^*$E-mail: mkv@iitk.ac.in}

\author{M. Chaudhuri}

\address{Department of Computer Science and Engineering, IIT Kanpur,\\
Kanpur, 208016, India}

\begin{abstract}
Pseudo-spectral method is one of the most accurate techniques for simulating turbulent flows.   Fast Fourier transform (FFT) is an integral part of this method.   In this paper, we present a new procedure to compute FFT in which we save operations during interprocess communications by avoiding transpose of the array.  As a result, our transpose-free FFT is 15\% to 20\% faster than FFTW.
\end{abstract}


\bodymatter

\section{Introduction}

A turbulent flow is chaotic. Hence, we need to employ accurate numerical schemes for simulating such flows.  A pseudospectral algorithm~\cite{Boyd:book} is one of the most accurate methods for solving fluid flows, and it is employed for  direct numerical simulations of turbulent flows, as well as for critical applications like weather prediction and climate modelling.   We have developed a general-purpose parallel flow solver named Tarang~\cite{Verma:Pramana2013,Tarang:web} for turbulence simulation.  At present, Tarang has solvers for incompressible flows involving pure fluid, Rayleigh B\'{e}nard convection, passive and active scalars, magnetohydrodynamics, liquid metals, etc. 

Fast Fourier transform (FFT) is the most important component of a psuedospectral solver, and it takes approximately 75\% of the total simulation time.  Hence, an efficient FFT is required for such solvers.  An FFT operation requires full data set, hence, it is one of the most difficult computational methods to parallelise.  The computational complexity of the algorithm is $O(N \log N)$, but it also involves \texttt{All\_to\_all} communications among the processors, which brings down the efficiency of the transform considerably. 

During an FFT operation, the processes need to exchange a large amount of data among themselves.  Most FFTs transpose the data during the exchange, and it takes around $15\%$ of the total data-exchange process.  In this paper, we present a new FFT scheme that does not perform transpose during the data exchange, thus provides a better performance. We also address the issue of {\em cache miss} during an FFT operations.

\section{FFT implementation}
We  perform FFT on a three-dimensional real array of size $n_0 \times n_1 \times n_2$, and transform it to a complex data of size $n_0 \times n_1 \times (n_2/2+1)$ and vice versa.  The real-to-complex (r2c) is called the forward transform, while the complex-to-real (c2r) is called the inverse transform.  In section 2.1, we will explain how FFT is performed in most of the present implementations using a transpose operations.  In Section 2.2, we introduce our transpose-free FFT.  
\begin{figure}[htbp]
    \centering
    \includegraphics[scale=.2]{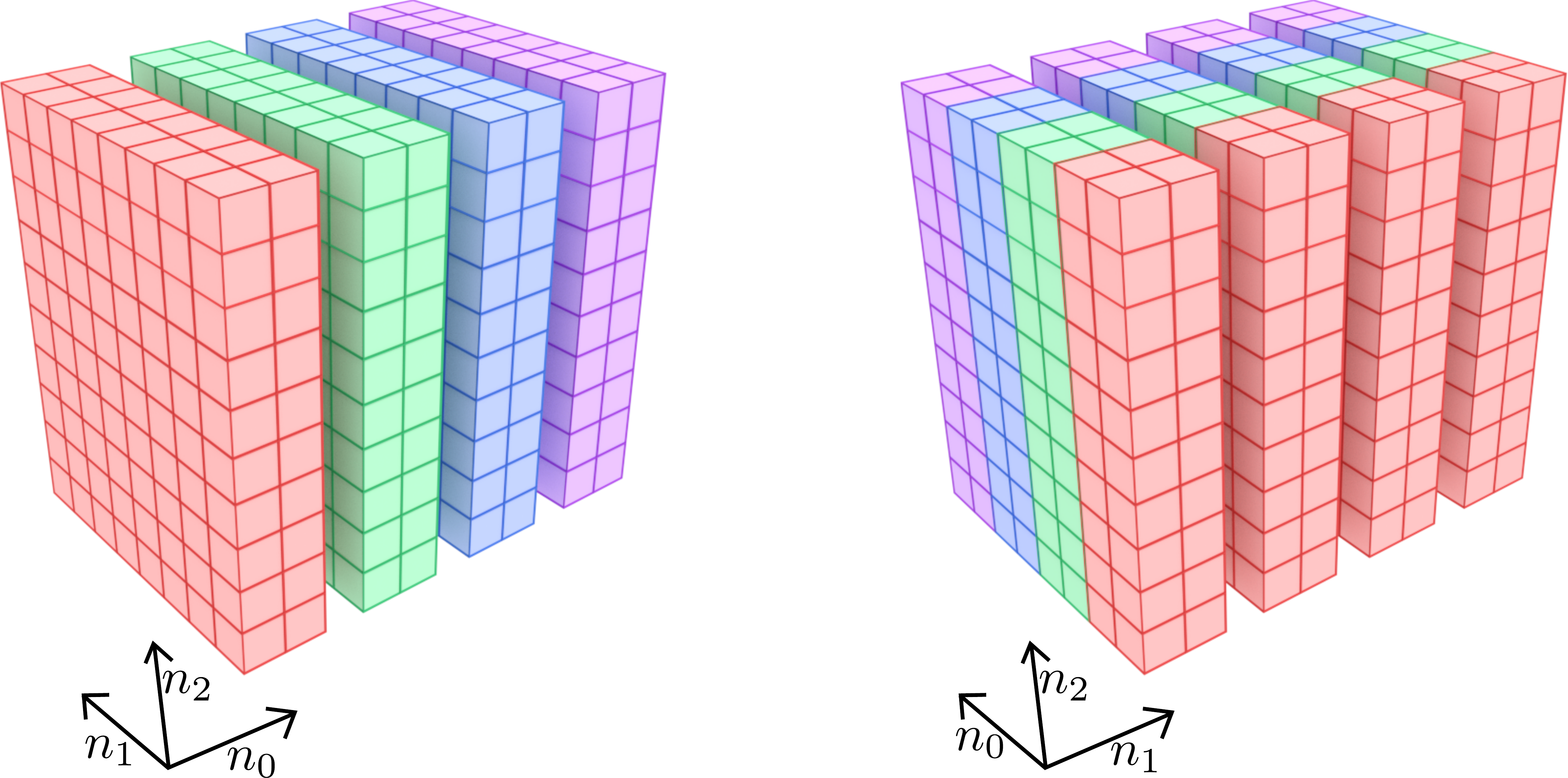}
    \caption{Data division $n_0 \times n_1 \times n_2$ array in FFT with transpose}
    \label{fig:3d_slab_FFTW}
\end{figure}

\subsection{FFT implementation with transpose}
For the parallel implementation of the forward transform, the real data is divided along the $n_0$ axis.  If there are $p$ processors, then each processor has $(n_0/p) \times n_1 \times n_2$ real data. After the forward transform, each processor has $(n_1/p) \times n_0 \times (n_2/2+1)$ complex data. A typical forward transform  involve three steps to be performed by each process:
\begin{enumerate}
	\item Two-dimensional forward transforms (r2c) of size $n_1 \times n_2$ for $n_0/p$ planes
	\item Transpose $n_0 \leftrightarrow n_1$, as shown in Fig. \ref{fig:3d_slab_FFTW}.
	\item One-dimensional c2c transform of size $n_0$ for $(n_1/p) \times (n_2/2+1)$ columns.
\end{enumerate}
We can  inverse transform  the complex data by performing the above operations in a reverse order.

\begin{figure}[htbp]
    \centering
    \includegraphics[scale=1]{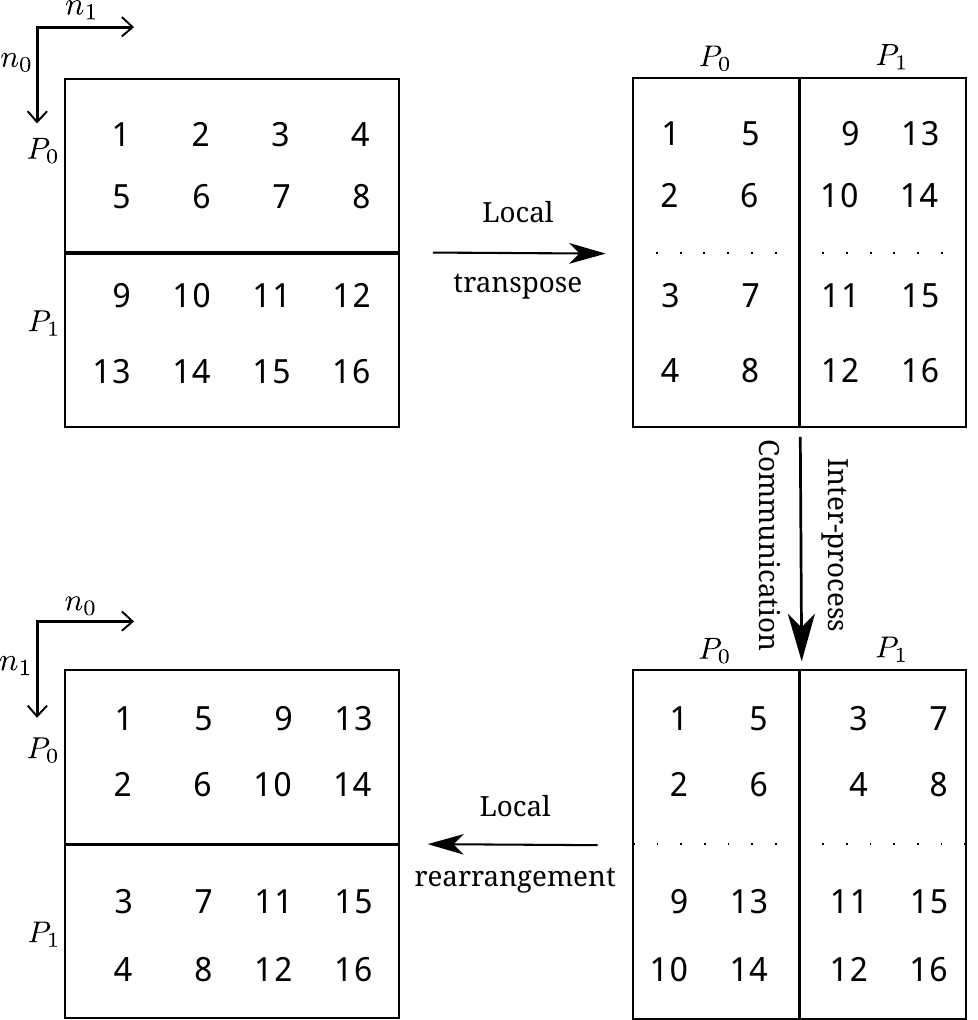}
    \caption{The standard transpose method}
    \label{fig:transpose_standard}
\end{figure}

The second step involves local transpose, interprocess communication, and again local rearrangement of data, as illustrated in Fig.~\ref{fig:transpose_standard} for a two-dimensional (2D) transform involving 16 data points and 2 processors.  Here  data along the $n_1$-axis is contiguous in memory and $n_0$-axis data is  spread in memory. 
Note that the axes $0$ and $1$ are flipped after the transpose.

Clearly, the data-exchange operation is quite expensive.  We devised a  scheme, somewhat similar to P3DFFT for pencil decomposition~\cite{Pekurovsky:JSciComput2012},  in which the data-exchange is less expensive.  We will describe our procedure in the next subsection.

\subsection{Transpose-free FFT}
\begin{figure}[htbp]
    \centering
    \includegraphics[scale=1]{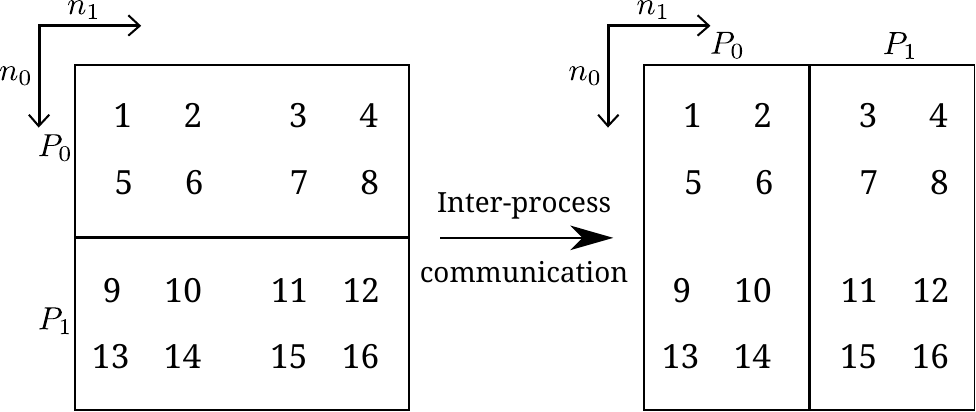}
    \caption{Data exchange using strided MPI functions, that does not require a local transpose}
    \label{fig:transpose}
\end{figure}

In our procedure, we modify step (2) of the inter-processor communication without any transpose.   In Fig.~\ref{fig:transpose}, we illustrate the communication process using a two-dimensional array.   We use  \texttt{MPI\_Type\_vector} and \texttt{MPI\_Type\_create\_resized} to select strided data, e.g., [3,4,7,8] for the process $P_0$, and [9,10,13,14] for the process $P_1$, and then send them to the other processor.  MPI functions \texttt{MPI\_Isend/MPI\_Recv} or \texttt{MPI\_All\_to\_all} are employed for communications; we choose the one  better-suited for a given hardware.  

\begin{figure}[htbp]
    \centering
    \includegraphics[scale=.2]{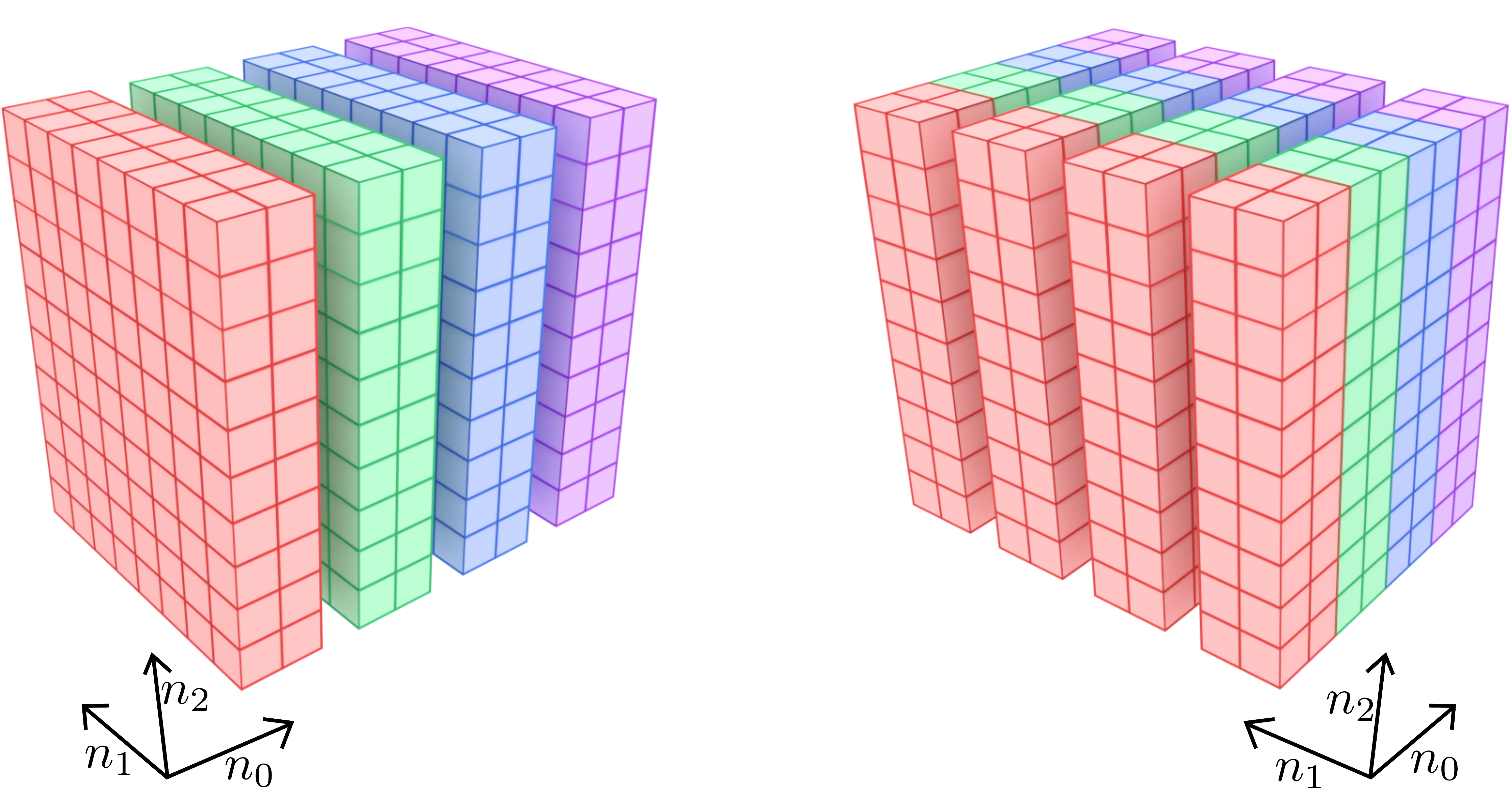}
    \caption{Data division $n_0 \times n_1 \times n_2$ array in transpose-free FFT}
    \label{fig:3d_slab_ST}
\end{figure}

The data structure after the interprocess communication is shown in Fig.~\ref{fig:3d_slab_ST}.  Here, the data along the $n_0$-axis for the third step are not contiguous, but strided. Therefore, we use strided FFTW functions  for this operation. 

Our procedure saves significant time during the data-exchange process.  The strided FFT function may appear to be more time-consuming, but that is not so because of the intelligent cache prefetch algorithms~\cite{Berg:UW2004}. Thus, our transpose-free FFT performs better than the ones with transpose.

\section{Comparison between the two procedures}

\begin{figure}[htbp]
    \centering
    \includegraphics[scale=.5]{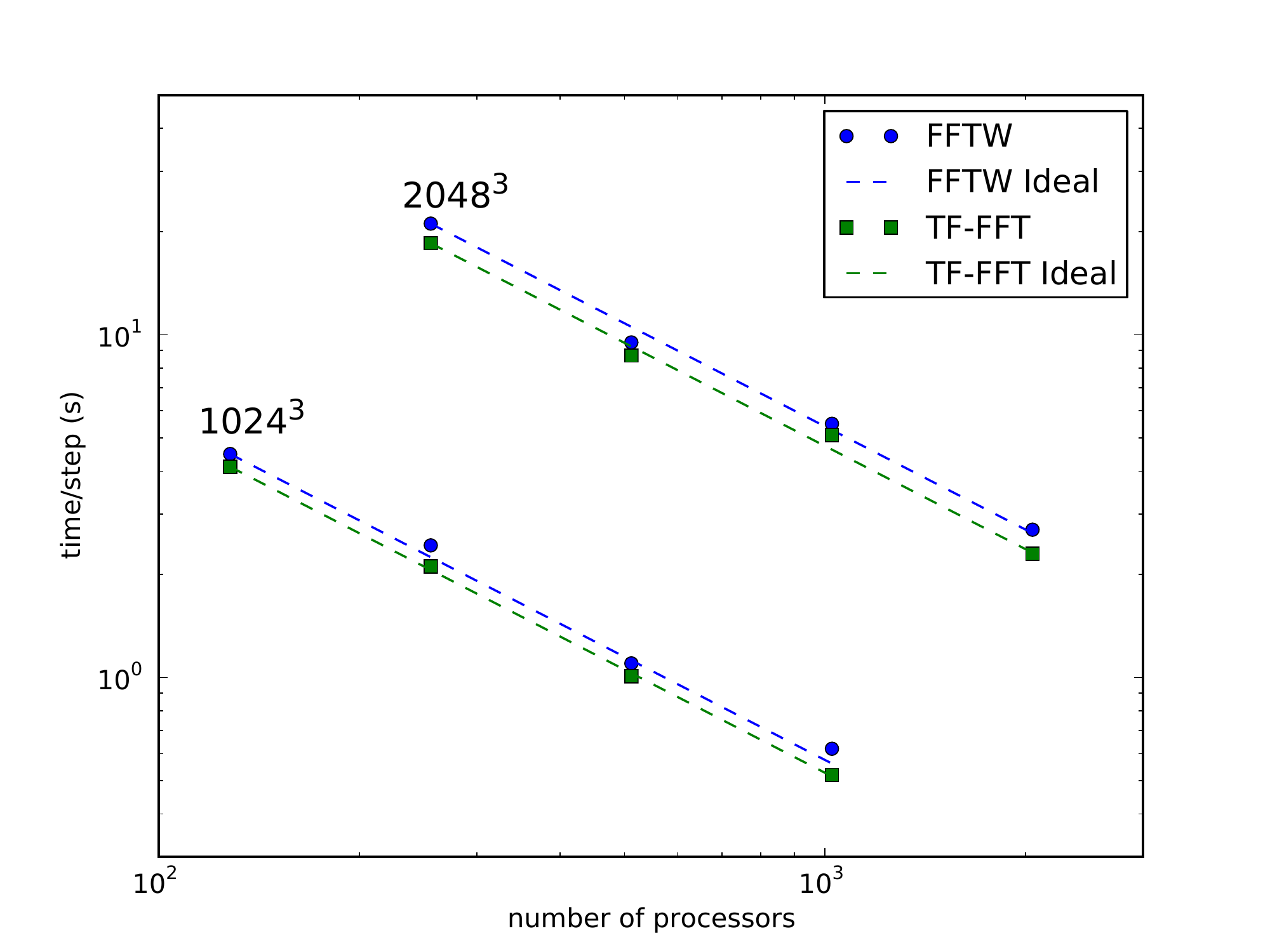}
    \caption{Comparison of Transpose Free FFT (TF-FFT) vs FFTW}
    \label{fig:st_vs_fftw}
\end{figure}

We compare the efficiencies of the two procedures by performing FFT (r2c and c2r) operations on $1024^3$ and $2048^3$ data sets.  We take the popular FFTW~\cite{FFTW:web} for FFT implementation with transpose. The maximum number of processor used in the comparison was  2048  of IBM Bluegene/P in KAUST.  The results are shown in Fig.~\ref{fig:st_vs_fftw}, in which the time taken by FFTW are represented by blue circles, while that by transpose-free FFT (named as TF-FFT) are shown by green squares.  We observe that the transpose-free FFT is  12\% to 16\% more efficient for $1024^3$ data, and 7.5\% to 14\% more efficient for $2048^3$ data. The gain in performance is due to the avoidance of multiple operations during step (2) of FFT with transpose (see Section 2.1).

We have incorporated the transpose-free FFT to Tarang 2.4, which yields approximately 2 times speed gain compared to Tarang 1.0~\cite{Verma:Pramana2013}.  Also, FFT is widely used in signal and image processing, data analysis, quantum mechanical computations; our proposed procedure would be useful for these applications as well.


\end{document}